\documentclass[aerospace,article,accept,pdftex,moreauthors]{Definitions/mdpi} 
\usepackage{xurl}
\firstpage{1} 
\pubvolume{1}
\issuenum{1}
\articlenumber{0}
\pubyear{2025}
\copyrightyear{2025}
\externaleditor{Firstname Lastname} % More than 1 editor, please add `` and '' before the last editor name
\datereceived{16 August 2025} 
\daterevised{15 September 2025} % Comment out if no revised date
\dateaccepted{15 September 2025} 
\datepublished{ } 
\hreflink{https://doi.org/} % If needed use \linebreak
\usepackage{savesym}

\savesymbol{tablenum}
\usepackage{siunitx}
\restoresymbol{SIX}{tablenum}
\usepackage{wrapfig}
\usepackage{bm}

\Title{Intercepting 3I/ATLAS at its %MDPI EE: Attention, title altered. Please check intended meaning has been retained.
	Closest Approach to Jupiter with the Juno Spacecraft}

\TitleCitation{Intercepting 3I/ATLAS at its Closest Approach to Jupiter with the Juno Spacecraft}

\Author{{Abraham Loeb} $^{1,}$*\orcidA{}, %MDPI: Please carefully check the accuracy of names and affiliations.
 Adam {Hibberd $^{2}$} %MDPI: The corresponding author is different from the one in the system. Please confirm whether this author is the corresponding author.
 and Adam Crowl $^{2}$}

\AuthorNames{Abraham Loeb, Adam Hibberd and Adam Crowl}

\isAPAStyle{%
       \AuthorCitation{Loeb, A., Hibberd, A., \& Crowl, A.}
         }{%
        \isChicagoStyle{%
        \AuthorCitation{Loeb, Abraham , Adam Hibberd, and Adam Crowl.}
        }{
        \AuthorCitation{Loeb, A.; Hibberd, A.; Crowl, A.}
        }
}

\address{%
$^{1}$ \quad Astronomy Department, Harvard University, 60 Garden Street, Cambridge, MA 02138, USA\\
$^{2}$ \quad Initiative for Interstellar Studies (i4is), 27/29 South Lambeth Road, London SW8 1SZ, UK; {adam.hibberd@i4is.org} (A.H.); {adam.crowl@i4is.org} (A.C.)}%MDPI: We added the email addresses here according to those submitted online at susy.mdpi.com. Please confirm.
\corres{Correspondence: {aloeb@cfa.harvard.edu}} %MDPI: The email addresses highlighted are different from the ones submitted online at susy.mdpi.com. Please confirm which are correct.
\abstract{\textls[-15]{The interstellar object 3I/ATLAS is expected to arrive at a distance of \mbox{$53.56(\pm 0.45)$~million~${\rm km}$}} ($0.358\pm 0.003$~au) from Jupiter on 16 March 2026. We show that applying a total thrust $\Delta$V of $2.6755~{{\rm km~s^{-1}}}$ to the lower perijove on 9 September 2025 and then executing a Jupiter Oberth Maneuver can bring the Juno spacecraft from its orbit around Jupiter to intercept the path of 3I/ATLAS on 14 March 2026. We further show that it is possible for Juno to come much closer to 3I/ATLAS ($\sim${27} million \si{km}) with 110 kg of remaining propellant, merely 5.4\% of the initial fuel reservoir. We find that for low available $\Delta$V, there is no particular benefit in the application of a double impulse (for example, to reach $\sim${27} million \si{km} from 3I/ATLAS); however, if Juno has a higher $\Delta$V capability, there is a significant advantage of a second impulse, typically saving propellant by a factor of a half. A close fly-by might allow us to probe the nature of 3I/ATLAS far better than telescopes on Earth.}

\keyword{3I/ATLAS; interstellar objects; Juno mission; trajectory}

\begin{document}

%%%%%%%%%%%%%%%%%%%%%%%%%%%%%%%%%%%%%%%%%%
\setcounter{section}{0} %% Remove this when starting to work on the~template.

% The order of the section titles is different for some journals. Please refer to the "Instructions for Authors” on the journal homepage.

\section{Introduction}
\label{sec1}

The interstellar object 3I/ATLAS {({Minor Planet Center}  %MDPI: 1. Footnotes are not allowed. We have moved the content of this section to the main text. Please confirm. 2. Hidden links are not allowed, we have displayed it, please confirm. 3. Please add the access date (format: Date Month Year), e.g., accessed on 1 January 2020. Same as below. 
% REPSONSE: DONE, A.H.
 (\url{https://www.minorplanetcenter.net/mpec/K25/K25N12.html}), accessed on 10 July 2025)}
was discovered on 1 July 2025 \citep{seligman2025discovery,Loeb_2025,bolin2025,alvarezcandal2025,opitom2025,Chandler2025,Belyakov2025}. It is expected {({NASA JPL SSD} \url{https://ssd.jpl.nasa.gov/tools/sbdb_lookup.html\#/?sstr=1004083}, accessed on 10 July 2025)} to arrive at a distance of $53.56(\pm 0.45)$~million~${\rm km}$ ($0.358\pm 0.003$~au) from Jupiter on {16 March 2026.} %MDPI: We have changed the dates in the article to "day month year" format. Please check the entire article.
% RESPONSE: DONE A.H.

{The `3I' designation indicates this is the third interstellar object to be discovered encountering our solar system. With~a heliocentric hyperbolic excess speed, \mbox{V$_{\infty}$ $\sim{58}$ \si{km.s^{-1}},} this object is unequivocally extrasolar and~joins the first interstellar object, 1I/'Oumuamua~\citep{Flekky2019,Seligman2020,Jackson2021,Desch2021,Bialy2018,Raymond2018}, and second, 2I/Borisov \citep{Jewitt_2019}, initially detected in 2017 and 2019, respectively, as~offering the potential of hitherto unparalleled scientific return on stellar systems far beyond our own in our Milky Way galaxy. They have conveniently provided us with an opportunity to study material from outside our solar system, without~actually sending an interstellar spacecraft. Such a craft  would otherwise take tens of thousands of years to arrive at its destination, using present day chemical rocket~technology.

This study focuses on the topic of sending an extant NASA spacecraft currently in orbit around Jupiter to intercept 3I/ATLAS. The~concept of sending spacecraft to an interstellar object is not new. Refer, for example, to the `Project Lyra' research for missions to 1I/$'$Oumuamua (undertaken largely by the Initiative for Interstellar Studies, i4is~\cite{I4IS}), which can be found in \citep{HPE19,HEL22,HHE20,HH21,AH23,HA23}{, and~also refer to~\cite{Seligman_2018}}. Furthermore, missions to 2I/Borisov have also been investigated by the i4is team \citep{HPH21}.

The analysis herein was conducted largely as a consequence of previous research which demonstrated that spacecraft missions to 3I/ATLAS from Earth are currently infeasible~\citep{yaginuma2025feasibilityspacecraftflybyinterstellar}. The~possibility of exploiting currently operating interplanetary spacecraft to observe 3I/ATLAS near its perihelion, when the Sun will shield it from Earth telescopes, has already been explored comprehensively in~\cite{eubanks20253iatlasc2025n1direct}. The~research expounded in the following is intended to elaborate on the potential capability of the NASA Juno probe, by~bringing it closer to the 3I/ATLAS object, when this object approaches Jupiter in March 2026.}
 
This close encounter provides a rare opportunity to shift the spacecraft Juno {({NASA Juno Mission} \url{https://www.jpl.nasa.gov/missions/juno/}, accessed on 10 July 2025)}
% RESPONSE: DONE A.H.
from its current orbit around Jupiter to intercept the path of 3I/ATLAS at its closest approach to Jupiter. The~instruments available on Juno, namely a near-infrared spectrometer, magnetometer, microwave radiometer, gravity science instrument, energetic particle detector, radio and plasma wave sensor, UV spectrograph and visible light camera/telescope, can all be used to probe the nature of 3I/ATLAS from a close~distance.

Below, we study the thrust required to shift Juno from its current orbit around Jupiter to a path that will intercept 3I/ATLAS in mid-March 2026.

\section{Orbit~Calculation}\label{sec2}

{There are two main strategies generally adopted for the determination of optimal trajectories (in this case, of spacecraft trajectories): the indirect method and the direct method \citep{CONWAY}. The~former normally involves expressing the problem in terms of a Hamiltonian and adjoint variables, and~then solving the trajectory, normally via trajectory integration, or~alternatively through collocation methods, and~using iterative steps until the initial adjoint variables are solved and the conditions for optimality, as~formulated by Pontyragin in his `Pontryagin Maximum Principle', including the target transversality conditions, are satisfied. This method ensures that the true optimal solution is found to be within the prescribed tolerance. The~application of this principle is wide in scope and can also be exploited to solve the problem of optimal N-impulse transfers \citep{jezewski1975primer}.

The alternative direct method \citep{CONWAY} normally involves simplifying the problem significantly by parameterizing the controls (such as assuming they evolve linearly with time) and finding the coefficients/control parameters which result in an extreme of the cost functional (such as minimizing fuel usage) over the course of the trajectory. The~method can be used in conjunction with a choice of global optimization paradigms involving iteration methods, such as Non-Linear Problem (NLP) solvers or Genetic Algorithms \citep{cage1994}, and so on. Depending on the parameterization adopted, this direct method may not find the precise theoretical solution, but~may come sufficiently close to an extent that it becomes indistinguishable from the theoretical solution.}

Our analysis  exploits the software package known as {{Optimum Interplanetary Trajectory Software}} %MDPI: Please confirm if the italics is unnecessary and can be removed. The following are the same.
% RESPONSE: ITALICS REMOVED A.H.
 (OITS). {Note that OITS employs a `direct method' strategy as explained above}. Further information regarding OITS is provided by~\cite{OITS_info,AH2} and~\cite{HPH21}. Two possible NLP solver options are available for the work conducted here, namely NOMAD \citep{LeDigabel2011} or \mbox{MIDACO~\citep{Schlueter_et_al_2009,Schlueter_Gerdts_2010,Schlueter_et_al_2013}.} This is a modified version of OITS in that the central body of interest is not the Sun but Jupiter. The~data for Juno is taken from the SPICE data website \citep{NAIF}, using the file {juno\_pred\_orbit.bsp}. {Furthermore, solar and planetary positions and velocities are derived from the SPICE data file {de430.bsp}.}

OITS solves the Lambert problem for one orbital cycle only: given two times $t_1$ and $t_2$, what are the two orbital arcs that connect them? Assuming that the positions at the beginning of the arc and the end of the arc are known, then there are two solutions, a~short way and a long way, equivalent to an angular sweep, $\theta$, and~the retrograde angular sweep, $2\pi - \theta$. Here, $\theta$ is found from the dot product of the initial and final position vector. Having determined the short way and long way solutions, the~way with the maximum $\Delta$V is rejected, leaving the desired, lowest $\Delta$V solution. This procedure is conducted iteratively with different trial values of $t_1$ and $t_2$ (within user-specified bounds), until~OITS has converged on the overall minimum $\Delta$V~solution. 

{In addition, in~order to model {Oberth Maneuvers} (refer~\cite{Blanco2021}), that is applications of $\Delta$V at points other than at an encounter with a celestial body (so normally at the periapsis of the central body), the~notion of {Intermediate Points} \citep{AH2} is utilized. An~Intermediate Point is a point on a spherical surface of user-specified radius from the center of attraction (such as, in~this case, Jupiter). The~longitude and latitude, $\theta$ and $\phi$, respectively, of~this specified radius are optimized by the NLP in question, along with the aforementioned times $t_1$ and $t_2$. This would normally necessitate a good guess by the user as to the exact value of this radial distance; however, for the version of OITS applied in this research, the~user may specify a range of distances, and~thus, this radial distance, $R$, can be optimized by the NLP also. Therefore, this allows full optimization of the point where the Oberth $\Delta$V is applied.}

To solve the Lambert problem, the~Universal Variable formulation is followed \citep{Bate1971}. We focus on an intercept (i.e., a fly-by) since a rendezvous, where the target's velocity is matched by the spacecraft, is out-of-the-question, owing to the excessively high hyperbolic speed of 3I/ATLAS relative to Jupiter ($\sim${65.9} \si{km.s^{-1}}).

The binary SPICE kernel file for the interstellar object 3I/ATLAS was also extracted from the NASA Horizons service, on 18 July 2025.

{The intercept distances derived in this research are around 0.36 au from Jupiter, which is almost precisely the radius of Jupiter’s Hill Sphere. As~far as Juno is concerned, it starts deep in Jupiter’s gravitational well, and~this is the overwhelming driver governing the magnitude of the probe’s intercept $\Delta$V. Thus, when simulations including the Sun’s influence were conducted, there was no evident difference in the required $\Delta$V, at~least discernible within the precision of the NLP software. Furthermore, $\Delta$V budgets are listed WITHOUT the provision for navigational errors and Mid-Course Corrections (MCCs). Such parameters are not within the scope of this research and are rather something that the NASA Juno team should determine with their `insider' knowledge of Juno's current status.}

{Note that the precise mass of residual propellant remaining in Juno's tanks, as~well as the current status of the probe's engine, is known only in full by the NASA Juno project team. Thus, the research herein is intended as a reference for the team to determine which options are the most appropriate given the probe's current condition. Note also that perturbations other than those mentioned above are not considered, as~they are negligible for the trajectories studied (at around 3 Jupiter radii) . For~instance, the~effect of Jupiter's J2 gravitational harmonic, which falls according to the inverse cube of Jupiter distance, would be insignificant. Additionally, Juno's orbit is considerably inclined to Jupiter's equator; thus, the influence of Jupiter's moons on Juno's orbit can be effectively discounted.}

{As this research amounts to a feasibility study, engine-specific parameters are not adopted, and~where applicable, estimated propellant requirements are provided with required $\Delta$V to allow NASA to determine itself the appropriate mission options to choose.}

{Despite the caveats mentioned above, for the REBOUND simulations (refer to later in this section), the~trajectories were also integrated with the Sun's perturbing gravitational force included, and~this was found to have no noticeable effect on the trajectory solutions.}

{Juno is hardened to survive close approaches to Jupiter. Thus, sensitive electronics are inside a radiation-shielded box.}

\begin{figure}[H]
%\hspace{-1.0cm}
\includegraphics[scale=0.23]{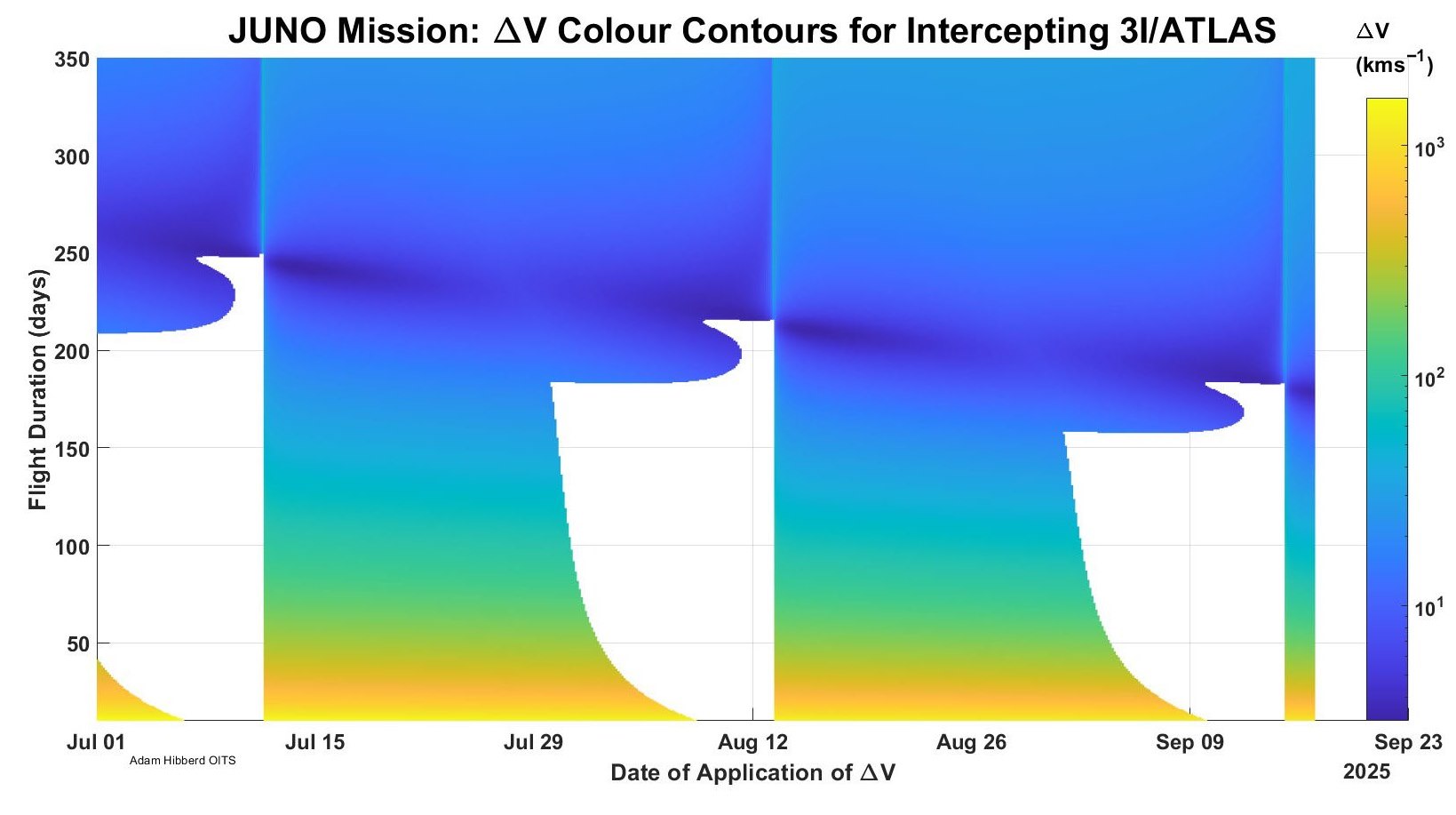}
\caption{Direct Juno Mission, {pork chop} %MDPI: 1. Please cite this figure in the text and ensure that the first citation of each figure appears in numerical order. 2. We revised the hyphen in the figure into minus sign. Please confirm.
% RESPONSE: FIGURES NOW CITED IN TEXT, AS REQUIRED, SWITCH TO MINUS CONFIRMED, A.H.
 indicating required $\Delta$V for the JUNO spacecraft to intercept 3I/ATLAS (logarithmic scale).}
\label{fig:CC1}
\end{figure}

\begin{figure}[H]
%\hspace{-1.0cm}
\includegraphics[scale=0.22]{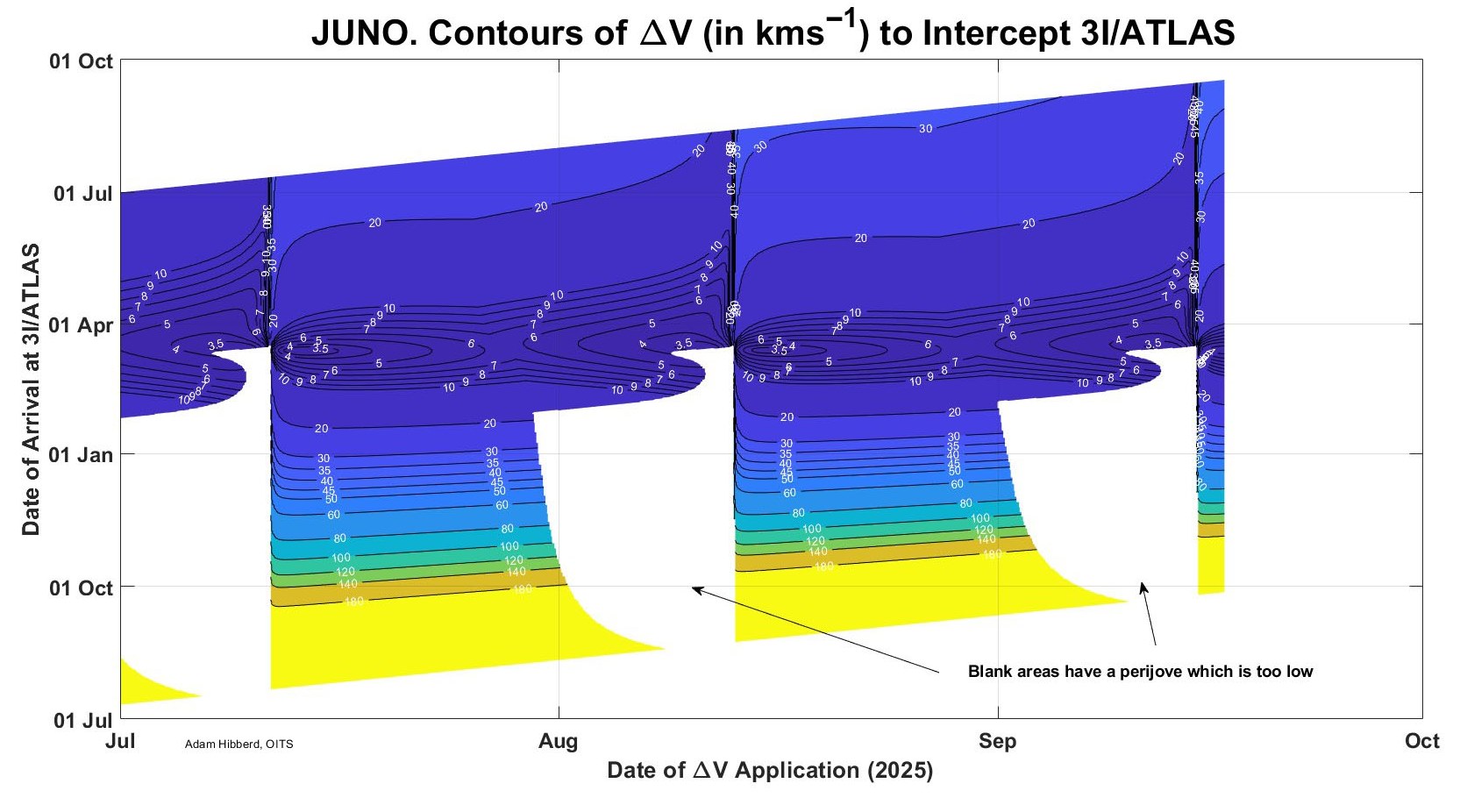}
\caption{Direct Juno Mission, {contours} %MDPI: 1. We revised the hyphen in the figure into minus sign. Please confirm. 2. Please confirm whether the numbers in the figure need additional explanation. 3. Please check whether the overlapping content in the figure affects the readability.
% RESPONSE: 1) FINE 2) WITH THE CHANGE IN CAPTION FOR CLARIFICATION 3) THE OVERLAPPING CONTENT IS IRRELEVANT AS IT IS OUTSIDE THE REGIONS OF INTEREST. A.H.
 of $\Delta$V needed by Juno to intercept 3I/ATLAS; note there are two opportunities before the binary SPICE kernel data ends on 17 September 2025.}
\label{fig:CC2}
\end{figure}

Using the approach outlined above, color contour maps were generated by OITS for a Juno $\Delta$V application window covering the present as at the time of writing (27 July 2025) to the point at which the data in the binary SPICE file for Juno expires (17 September 2025), marking the possible end of the mission which is currently scheduled to occur around that~time. Refer to Figures \ref{fig:CC1} and \ref{fig:CC2}.

The feasibility of intercepting 3I/ATLAS depends on the current amount of fuel available from the propulsion system of Juno. However, some inferences can be drawn from the total $\Delta$V available at the beginning of the Juno mission. On~its interplanetary trajectory, Juno conducted two Deep Space Maneuvers (DSMs), and~one Jupiter orbital insertion, both of which would have placed a significant demand on the chemical propulsion employed by Juno (Hydrazine and oxidizer nitrogen tetroxide).

Let us assume a total initial wet mass of the spacecraft $M_{tot}$, a~dry mass of $M_{dry}$, and~a specific impulse given by $I_{sp}$; then, the total $\Delta$V available to Juno is given by
\begin{equation}
\Delta V = I_{sp}\ g\ ln \left( \frac{M_{tot}}{M_{dry}} \right)
\label{Isp}
\end{equation}
where $g =9.8$~\si{m.s^{-2}}.

The data for the above parameters can be sourced {from} \cite{JUNOPK}. %MDPI: Please confirm which reference this mention is. Please make sure that the References are mentioned in the order.
% RESPONSE: PLEASE REFER TO FOLLOWING FILE
%@misc{JUNOPK,
%title="{Juno Launch Press Kit}",
%author = "{NASA}",
%year = {2011},
%url = {https://www.jpl.nasa.gov/news/press_kits/JunoLaunch.pdf}
%}

 Thus, we have \mbox{$M_{tot} = 3625$ \si{kg}} and $M_{dry} = 1593$  \si{kg}. For~the specific impulse, we assume an optimistic $I_{sp} = 340$ \si{s}, giving an overall initial $\Delta$V
available of 2.74 \si{km.s^{-1}}.

This value is similar to the required $\Delta$V for Juno to intercept 3I/ATLAS, given in Tables~\ref{tab1}--\ref{SOLOB}.

\begin{table}[H]\setlength{\tabcolsep}{0.9mm}
\caption{Pertinent trajectory data for an intercept of 3I/ATLAS with $\Delta$V applied in mid-August; refer to the middle trough in {Figure}~\ref{fig:CC2}. %MDPI: Figure 2 is mentioned before Figure 1. Figure 1 is not mentioned. Please revise it. Please make sure the Figures are mentioned in the order mentioned.
% RESPONSE, Figure 1 & Figure 2 NOW REFERENCED IN TEXT AT HEAD OF PAGE 4 AS REQUIRED, A.H.
}
\label{tab1}
%\resizebox{\textwidth}{!}{%
\small
\begin{adjustwidth}{-\extralength}{0cm}
\centering %% If there is a figure in wide page, please release command \centering
\begin{tabular}{cccccccc}
\toprule
\textbf{Number} & \textbf{Event}  & \textbf{Time}                   & \textbf{Arrival Speed} & \textbf{Departure Speed} & \textbf{\boldmath{$\Delta$}V} & \textbf{Distance from Jupiter} & \textbf{Perijove Alt.}\\ 
       &         &                        & \textbf{m/s}           & \textbf{m/s }            & \textbf{m/s}    & \textbf{km}                    & \textbf{km}       \\ \midrule
1      & Juno    & 2025 AUG 11 03:19:36   & 0             & 3259.3          & 3259.3 & 2,303,610               & 63,276    \\ \midrule
2      & 3I/ATLAS & 2026 MAR   16 01:32:30 & 66,129.2       & 66,129.2               & 0      & 53,392,590              & 63,276    \\ \midrule
 &  & & & Contingency Margin & 490 (15\%) & & \\ 
 &  & & & \textbf{{Total} \boldmath{$\Delta$}V} %MDPI: Please confirm if the bold is unnecessary and can be removed. The following are the same.
 % RESPONSE, BOLD CONFIRMED AS NECESSARY, A.H.
 & \textbf{{3749.3}} & & \\ \bottomrule
\end{tabular}%}

\end{adjustwidth}
\end{table}
\unskip

\begin{table}[H]\setlength{\tabcolsep}{1mm}

\caption{Trajectory data for an intercept of 3I/ATLAS with $\Delta$V applied in mid-September; refer to the~trough on the right in Figure~\ref{fig:CC2}. }
\label{tab2}
\small
%\resizebox{\textwidth}{!}{%

\begin{adjustwidth}{-\extralength}{0cm}
%\centering %% If there is a figure in wide page, please release command \centering
\begin{tabular}{cccccccc}
\toprule
\textbf{Number} & \textbf{Event} & \textbf{Time}                   & \textbf{Arrival Speed} & \textbf{Departure Speed} & \textbf{\boldmath{$\Delta$}V} & \textbf{Distance from Jupiter }& \textbf{Perijove Alt.}\\ 
       &         &                        & \textbf{m/s}           & \textbf{m/s }            & \textbf{m/s}    & \textbf{km}               & \textbf{km}       \\ \midrule
1      & Juno    & 2025 SEP 12 22:29:49   & 0             & 3306.5          & 3306.5 & 2,235,639           & 60,390    \\ \midrule
2      & 3I/ATLAS & 2026 MAR   16 11:45:48 & 66,068.9       & 66,068.9            & 0      & 53,331,939          & 60,390    \\ \midrule
 &  & & & Contingency Margin & 500 (15\%) & & \\ 
 &  & & & \textbf{{Total} \boldmath{$\Delta$}V} & \textbf{{3806.5}} %MDPI: 
 & & \\ \bottomrule
\end{tabular}%}
\end{adjustwidth}

\end{table}
\unskip

\begin{table}[H]\setlength{\tabcolsep}{0.4mm}
\caption{Jupiter Oberth Maneuver offers a lower $\Delta$V requirement than the direct~option.}
\label{SOLOB}
%\resizebox{\textwidth}{!}{%
\small
\begin{adjustwidth}{-\extralength}{0cm}
\centering %% If there is a figure in wide page, please release command \centering
\begin{tabular}{cccccccc}
\toprule
\textbf{Number} & \textbf{Event}             & \textbf{Time}                 & \textbf{Arrival Speed }& \textbf{Departure Speed} & \textbf{\boldmath{$\Delta$}V} & \textbf{Distance from Jupiter} & \textbf{Perijove Alt.}     \\
       &                    &                      & \textbf{m/s}           & \textbf{m/s}             & \textbf{m/s}    & \textbf{km} & \textbf{km} \\ \midrule
1      & Juno               & 2025 SEP 09 22:40:01 & 0             & 2157.4          & 2157.4 &          3,863,491 &     \\ \midrule
2      & 2.68 Jupiter Radii & 2025 SEP 14 18:37:14 & 35,881.8       & 36,388.6         & 518.1  & 191,595 & 120,103          \\ \midrule
3      & 3I/ATLAS           & 2026 MAR 14 12:51:04 & 66,536.8       & 66,536.8               & 0 & 54,576,427     & 88,660       \\ \midrule
 &  & & & Contingency Margin & 400 (15\%) & & \\ 
 &  & & & \textbf{{Total} \boldmath{$\Delta$}V} & \textbf{{3075.5}} %MDPI: 
 & & \\ \bottomrule
\end{tabular}%}
\end{adjustwidth}
\end{table}
\unskip
\unskip

\begin{figure}[H]
%\centering
\includegraphics[scale=0.27]{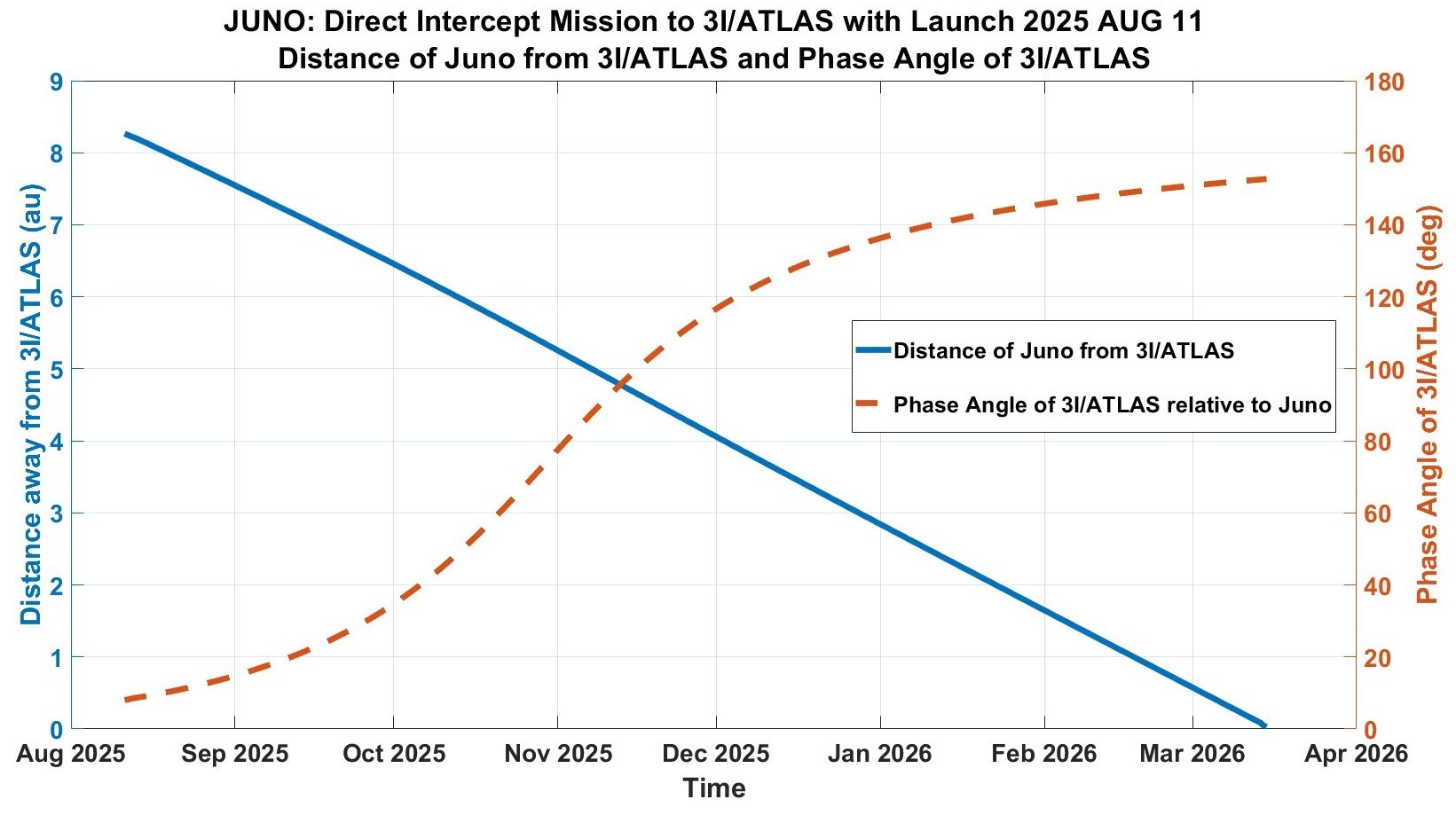}
\caption{Distance of 3I/ATLAS from Juno and Phase of 3I/ATLAS w.r.t. Juno, for~the reference mission (direct transfer) with a ‘launch' on 11 August 2025 (refer text).}
\label{fig:DistPhase}
\end{figure}

\begin{figure}[H]
%\centering
\includegraphics[scale=0.2]{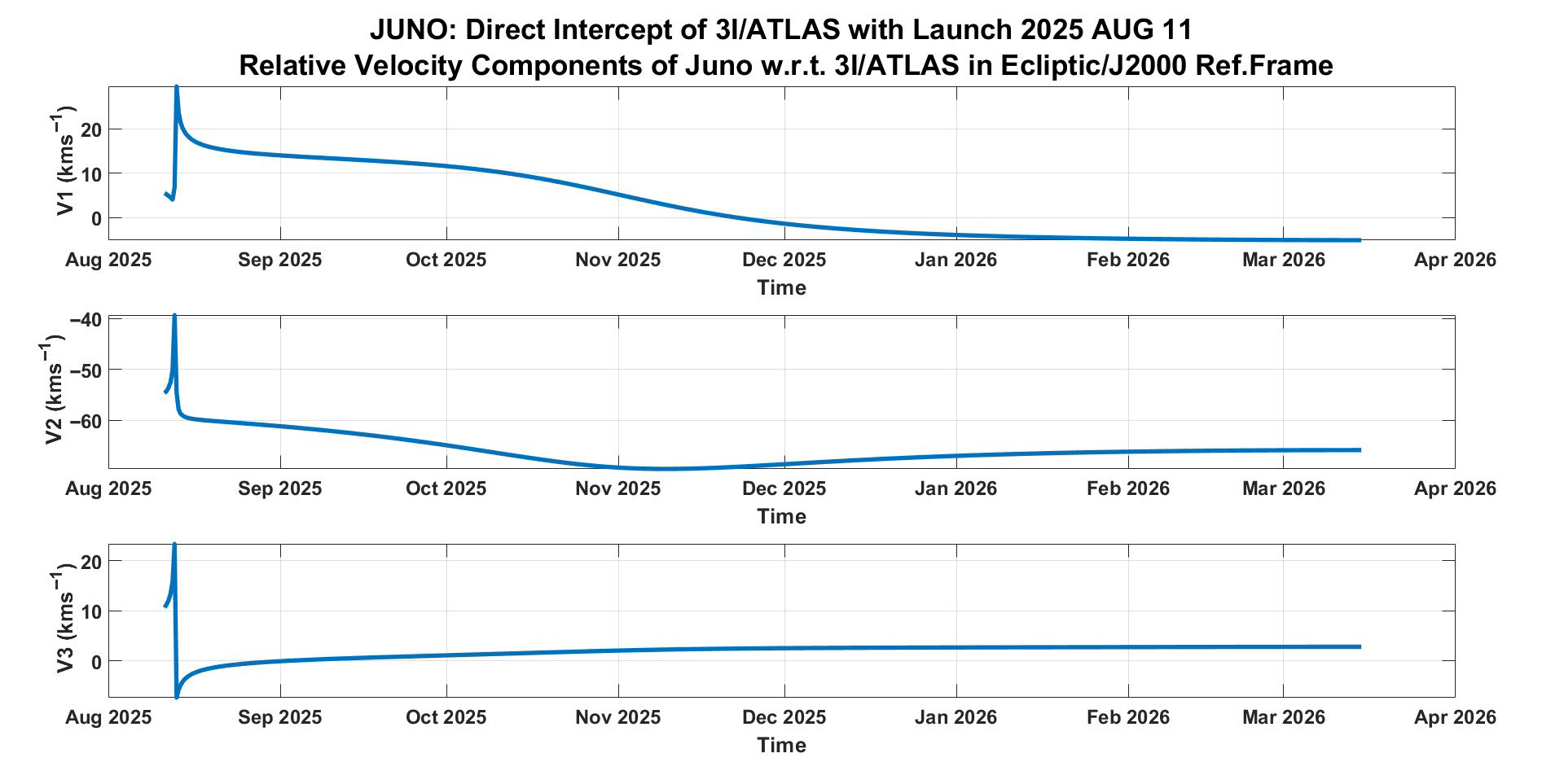}
\caption{{Reference} %MDPI: We revised the hyphen in the figure into minus sign. Please confirm.
% RESPONSE: FINE A.H.
 mission: relative velocity components of Juno w.r.t. 3I/ATLAS in the NASA SPICE ECLIPJ2000 reference~frame.}
\label{fig:Vrel}
\end{figure}

\begin{figure}[H]
%\centering
\includegraphics[scale=0.2]{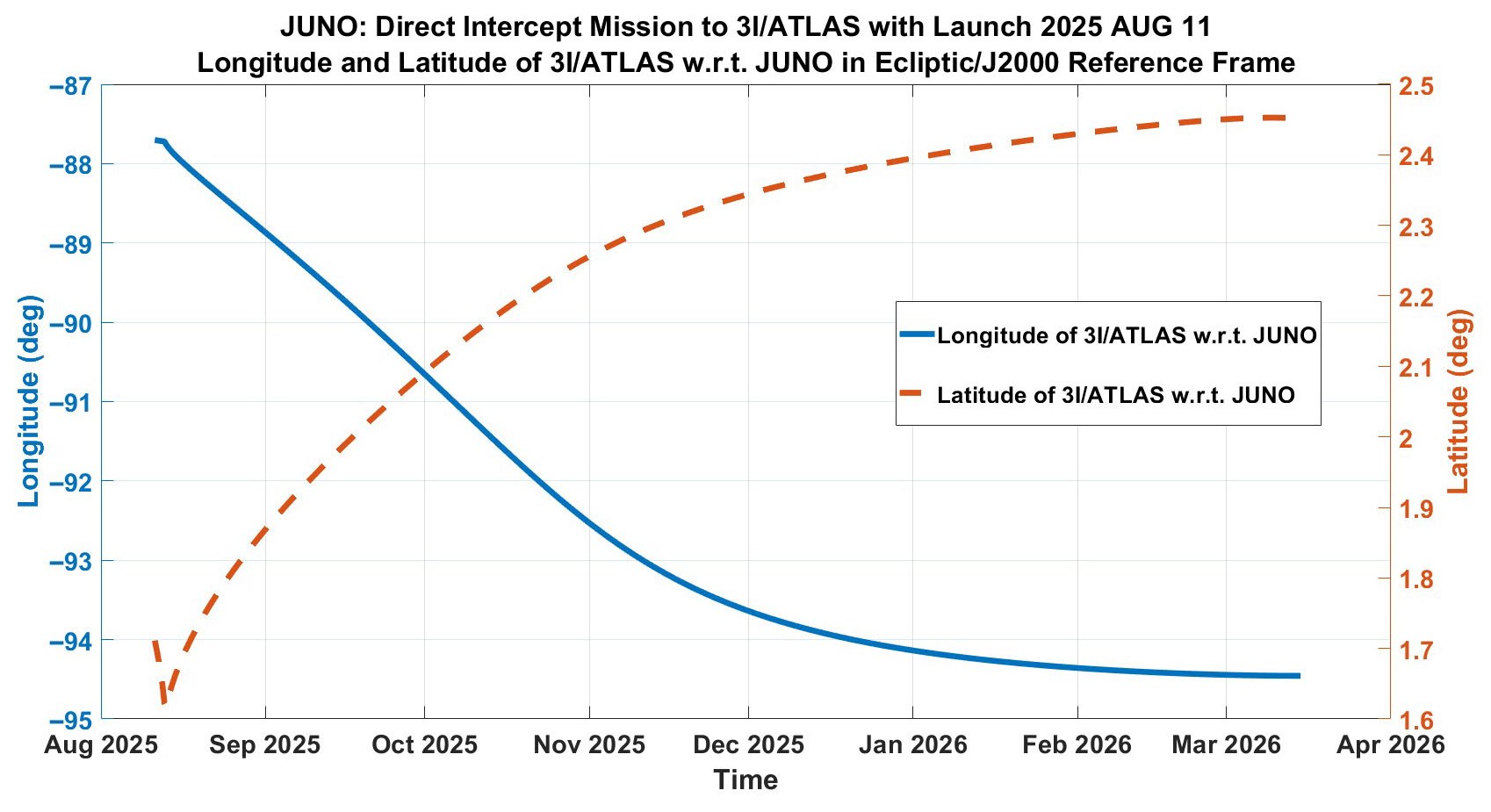}
\caption{{Longitude} %MDPI: We revised the hyphen in the figure into minus sign. Please confirm.
% RESPONSE: FINE A.H.
 and latitude of the vector from Juno to 3I/ATLAS in the NASA SPICE ECLIPJ2000 for attitude tracking of the target for reference~mission}
\label{fig:Attit}
\end{figure}

\begin{figure}[H]
%\centering
\includegraphics[scale=0.3]{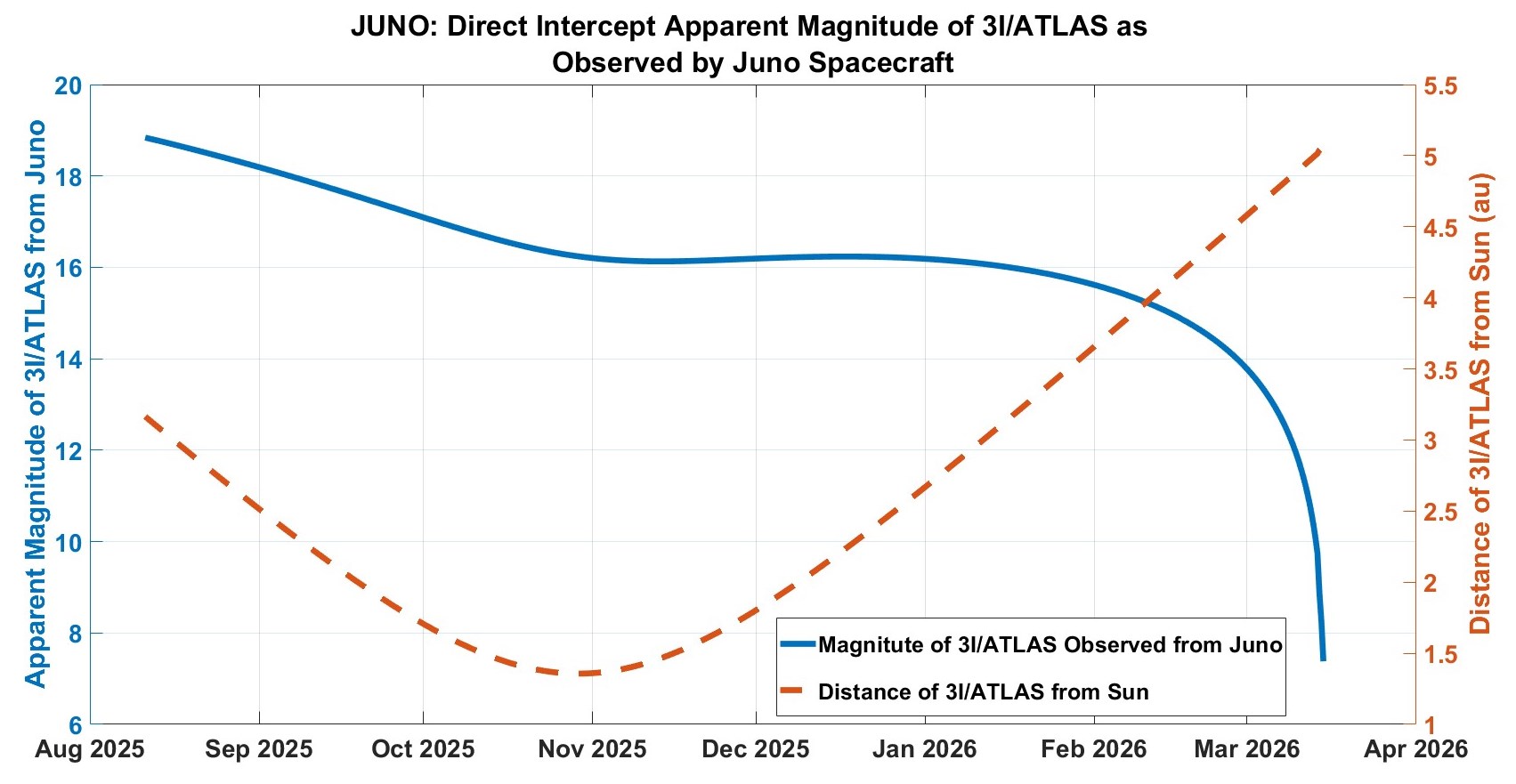}
\caption{Estimated apparent brightness of 3I/ATLAS w.r.t. Juno (left axis) and distance of 3I/ATLAS from Sun (right axis) for reference~mission. }
\label{fig:Mag}
\end{figure}

{The opportunity shown in Table~\ref{tab1} has a `launch' (henceforth defined as time of initial $\Delta$V application) of 11 August 2025, and~in what follows, this will be the reference mission. Refer to Figures~\ref{fig:DistPhase}--\ref{fig:Mag}, which provide the pertinent trajectory data for this reference mission, including an estimate of the brightness of 3I/ATLAS, as~well as the necessary attitude of Juno for tracking the target. Further missions investigated follow similar profiles and so are not provided.}

\unskip
\unskip
\unskip

Although the engine of Juno was not operated since 2016, the~required $\Delta$V might potentially be within Juno's performance envelope. In~that case, Juno would be able to get close  to 3I/ATLAS and use its instruments to probe the nature of the interstellar object and any cloud of gas or dust around~it.

The optimal option involves a Jupiter Oberth Maneuver which requires an application of $\Delta$V on 9 September 2025, only 8 days prior to the originally intended termination date for Juno's plunge into the atmosphere of Jupiter.
Having delivered this thrust to diminish Juno's altitude, a~further $\Delta$V is subsequently delivered, constituting a Jupiter Oberth Maneuver and resulting in an eventual intercept of the target 3I/ATLAS on 14 March 2026. Refer to Table~\ref{SOLOB} for more details. In~total, an~overall \mbox{$\Delta$V of $(2.1574 + 0.5181) = 2.6755$ \si{km.s^{-1}}} is~utilized. 

If doable, this exciting new goal will rejuvenate Juno’s mission and extend its scientific lifespan beyond 14 March 2026.

So far, we have examined a zero distance intercept of Juno with 3I/ATLAS. It is salient at this juncture to ask the question ``how close can Juno approach 3I/ATLAS, given that it has a limited remaining propellant mass, and~so a restricted $\Delta$V?''.

In order to perform these investigations, a~software application using SPICE \citep{NAIF}, REBOUND \citep{2012A&A...537A.128R,2015MNRAS.446.1424R} and NOMAD \citep{LeDigabel2011} was~constructed.

\begin{figure}[H]
%\centering
\includegraphics[scale=0.5]{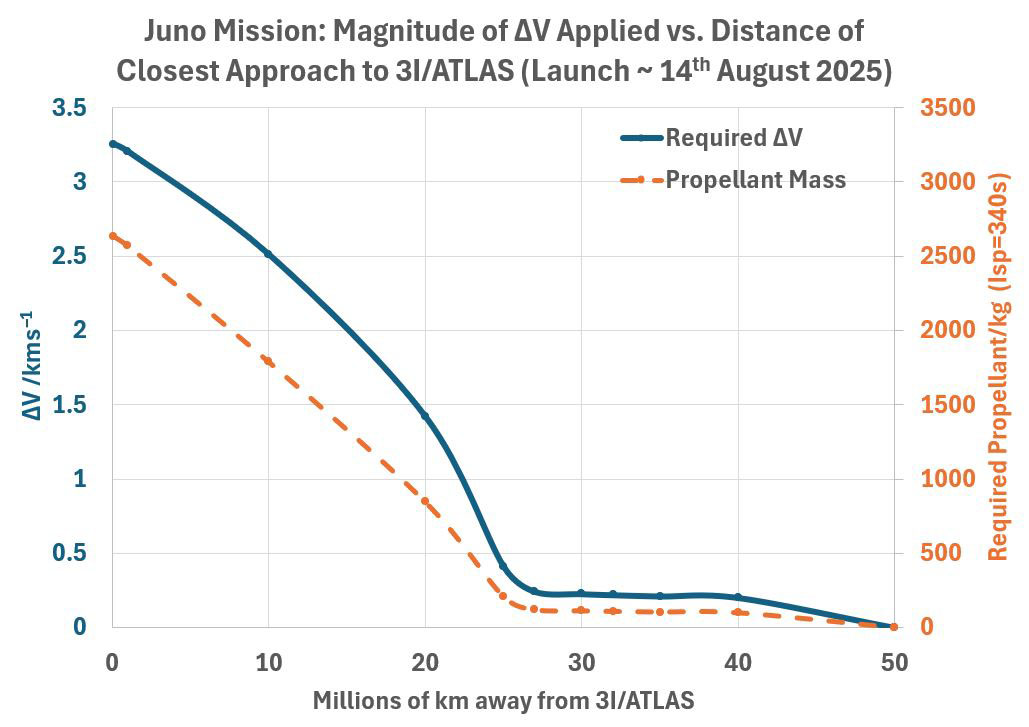}
\caption{{Thrust} %MDPI: We revised the hyphen in the figure into minus sign. Please confirm.
% RESPONSE: FINE A.H.
 impulse $\Delta$V (left vertical axis) and propellant mass (right vertical axis) needed for Juno to come within a range of distances from 3I/ATLAS (horizontal axis). The~launch date is assumed to be 14 August 2025.}
\label{fig:Dist_August}
\end{figure}

We investigate the mid-August 2025 opportunity first. Figure~\ref{fig:Dist_August} assumes a rocket specific impulse, $I_{sp} = 340$s, and uses Equation~(\ref{Isp}) to derive the
required propellant mass for a given $\Delta$V. We find that a relatively low $\Delta$V is needed (<0.23 \si{km.s^{-1}}, equivalent to a propellant mass of $\sim${110} \si{kg}, which is merely 5.4\% of the initial fuel reservoir) to approach 3I/ATLAS within a distance of 27 million km. Below~$\sim$27 million km, the~required $\Delta$V rises significantly until it reaches 3.3 \si{km.s^{-1}} at zero distance as determined in the preceding analysis results, presented in Table~\ref{tab1}.

\begin{figure}[H]
%\centering
\includegraphics[scale=0.5]{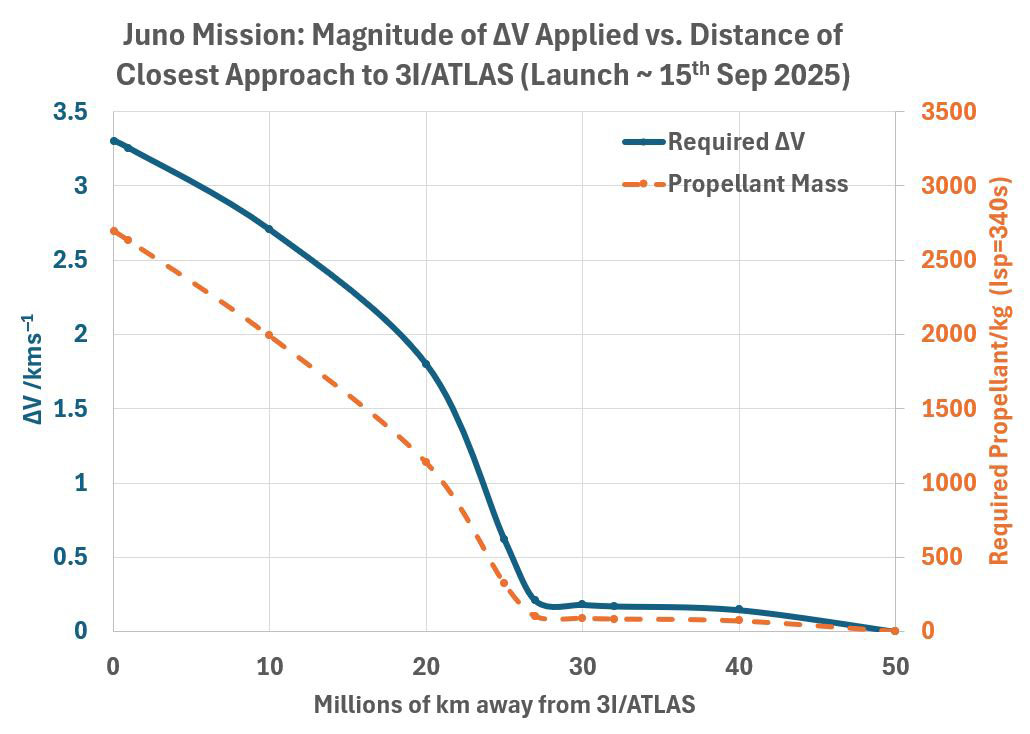}
\caption{{Thrust} %MDPI: We revised the hyphen in the figure into minus sign. Please confirm.
% RESPONSE: FINE A.H.
 impulse $\Delta$V (left vertical axis) and propellant mass (right vertical axis) needed for Juno to come within a range of distances from 3I/ATLAS (horizontal axis). The launch date is assumed to be 15 September 2025.}
\label{fig:Dist_Sep}
\end{figure}

Figure~\ref{fig:Dist_Sep} refers to the September 2025 opportunity and shows a similar behavior  with similar levels of required $\Delta$V and propellant mass, though~the advantage of this option is that it provides a month of extra time to prepare for the~maneuver.

\unskip

\begin{figure}[H]
%\centering
\includegraphics[scale=0.47]{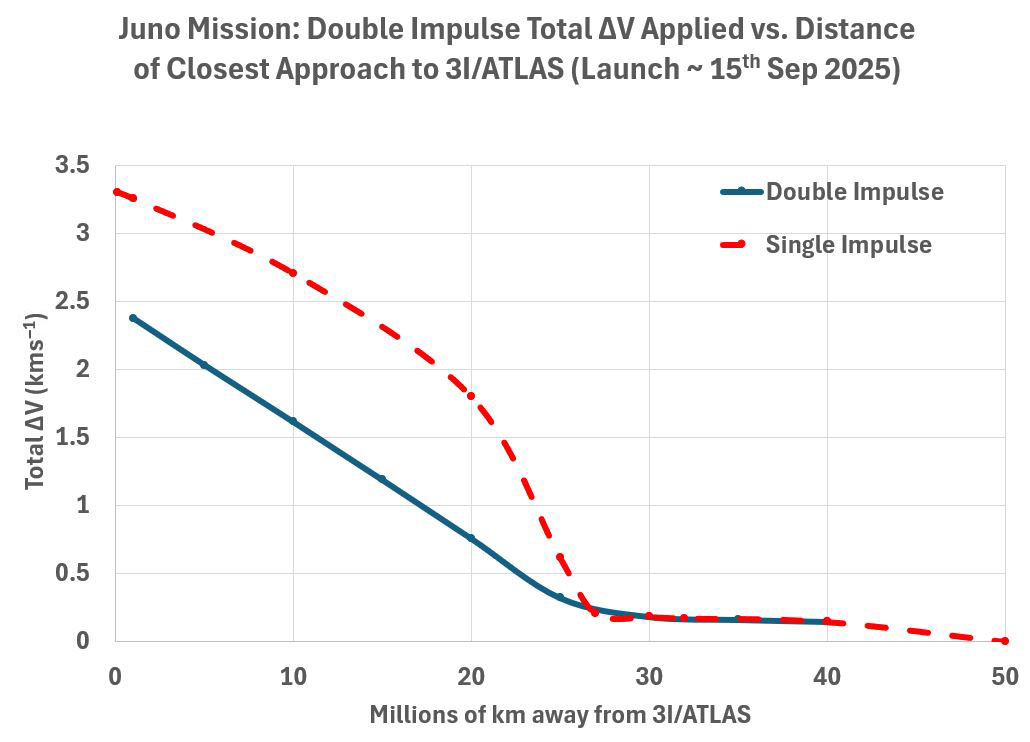}
\caption{{Double} %MDPI: We revised the hyphen in the figure into minus sign. Please confirm.
% RESPONSE: FINE A.H.
 impulse scenario (assuming the opportunity around 15 September 2025) compared to the single impulse option, implying a significant reduction in required total $\Delta$V.}
\label{fig:Dist_Sep2}
\end{figure}

\begin{figure}[H]
%\centering
\includegraphics[scale=0.47]{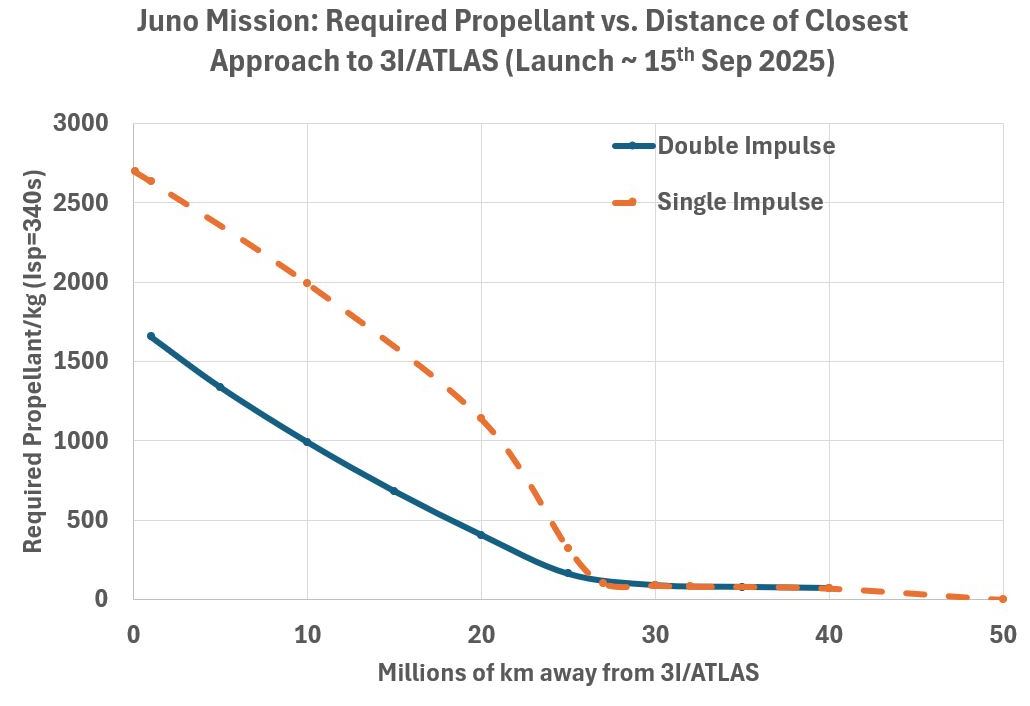}
\caption{Double impulse scenario compared with the single impulse option, as~for Figure~\ref{fig:Dist_Sep2}, in~terms of required propellant~mass.}
\label{fig:Fuel_Sep2}
\end{figure}

There would potentially be a lower overall $\Delta$V requirement with more than one impulse application. For~simplicity, we investigate here only the double impulse case and focus on the mid-September 2025 launch. Figure~\ref{fig:Dist_Sep2} compares the $\Delta$Vs of the double impulse with the single impulse option. There is a significant drop in the required $\Delta$V for the double impulse scenario. Figure~\ref{fig:Fuel_Sep2} shows how this $\Delta$V translates to required propellant, implying a factor of a half reduction in the propellant mass needed in order to get to a distance of \mbox{10 million km} from 3I/ATLAS.

\begin{figure}[H]
%\centering
\includegraphics[scale=0.5]{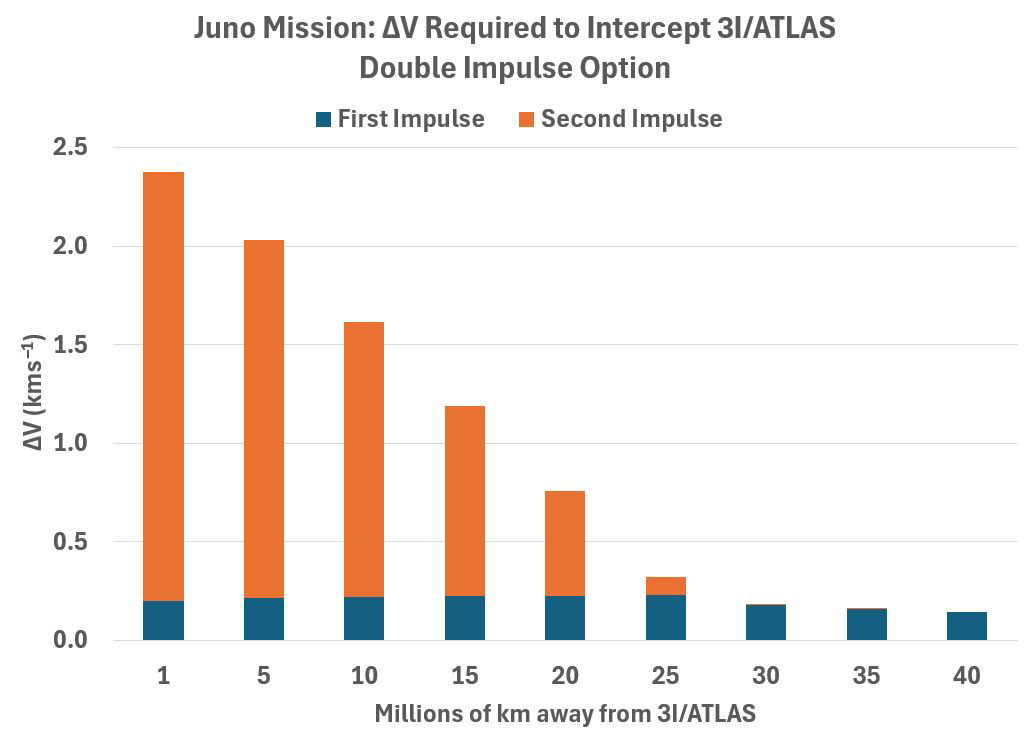}
\caption{{Double} %MDPI: We revised the hyphen in the figure into minus sign. Please confirm.
% RESPONSE: FINE A.H.
 impulse scenario: how the $\Delta$V is distributed between the 1{st} impulse (blue section) and 2{nd} impulse (red section).}
\label{fig:Ratio_Sep2}
\end{figure}

Figure~\ref{fig:Ratio_Sep2} shows the distribution of $\Delta$V between the 1{st} impulse (blue section) and 2{nd} impulse (red section), indicating that for the situation where only small $\Delta$Vs are available (i.e., closest approach to 3I/ATLAS $>$ 25 million \si{km}), there is no benefit at all to choosing the additional impulse. We note that the double impulse is more challenging to realize given the limited time remaining for~preparation.

\vspace{-6pt}
\unskip
\unskip

\begin{figure}[H]
%\centering
\includegraphics[scale=0.5]{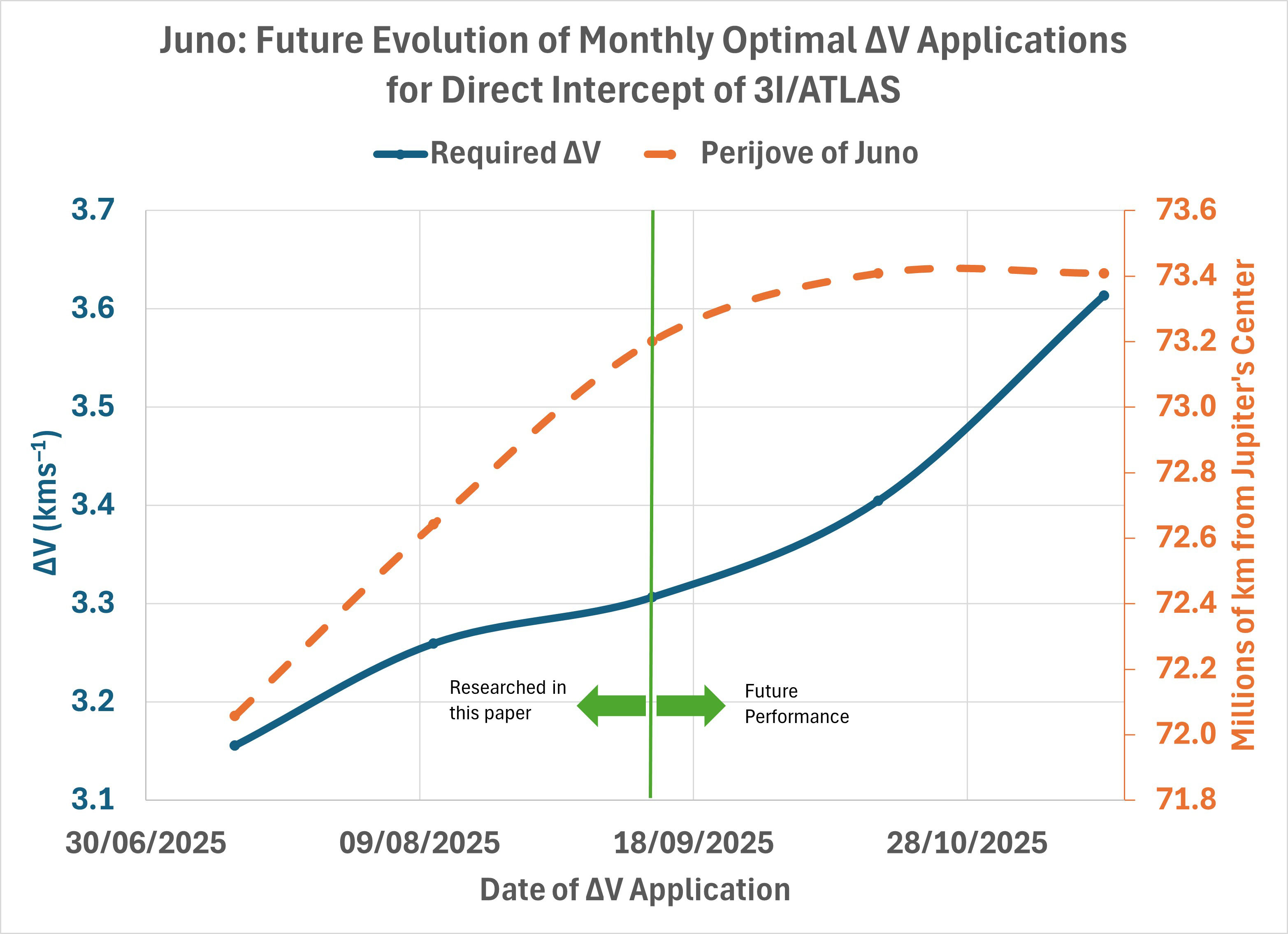}
\caption{{Evolution} %MDPI: We revised the hyphen in the figure into minus sign. Please confirm.
% RESPONSE: FINE A.H.
 of $\Delta$V (dark solid line) needed for the direct intercept scenario studied in Section~\ref{sec2} is provided beyond September 2025 and up to November 2025. The~perijove of Juno (red dashed line) as it evolves with time is also provided as this influences the level of $\Delta$V needed from the Juno~probe. }
\label{fig:Future}
\end{figure}

{
\section{Future~Opportunities}

The investigations in the preceding section were conducted up to the end of the NASA SPICE kernel file for Juno, which is in late September 2025, and~so they ignore any opportunities which may arise after this cut-off point. Since more up-to-date data is now available from NAIF, extending the Juno ephemerides to late November 2025, this offers the opportunity to characterize the future evolution of the performance of a Juno mission to 3I/ATLAS.

Thus, refer to Figure~\ref{fig:Future}, which extends the level of $\Delta$V needed for a direct intercept beyond August and September (refer to Tables~\ref{tab1} and \ref{tab2}, respectively) and up to the end of November. This provides a good idea of how the other trajectory scenarios investigated in Section~\ref{sec2} will perform in the~future.

Evidently, there are two key factors which govern the level of $\Delta$V required to intercept 3I/ATLAS:

\begin{enumerate}
    \item The time available from application of the initial $\Delta$V to reach and intercept the target, 3I/ATLAS.
    \item The precise perijove of Juno, since the lower this value, the~greater the kick which can be delivered from the Oberth effect using the same magnitude $\Delta$V.
\end{enumerate}

As can be observed in Figure~\ref{fig:Future}, the~perijove of Juno ascended from July onwards and then plateaued in October/November. This implies that in the future, from~November onwards, there will be no obvious benefit from Juno's orbit in terms of reducing the applied $\Delta$V to arrive at 3I/ATLAS. In~fact, if one refers to the plot of required $\Delta$V in this Figure, we find that this important metric of feasibility continues quite steeply upwards despite this plateau in perijove, implying that it is indeed the first of the items above (lower flight durations due to the more imminent arrival of 3I/ATLAS) that is the overriding factor governing the viability of a mission to 3I/ATLAS. In~short, the sooner Juno is `launched' the better.
}
\unskip

\section{Discussion}

{The amount of fuel left in Juno’s engine is not publicly available. As~a result, our paper provided the distance of the closest approach that Juno can reach relative to 3I/ATLAS as a function of that unknown fuel reservoir. This is the best that can be done at this time, following email exchanges we had with the Juno team.
 }

We have found that the application of a thrust of $2.6755~{\rm km~s^{-1}}$ on 9 September 2025 can potentially shift the Juno spacecraft from its orbit around Jupiter to intercept the path of 3I/ATLAS on 14 March 2026.

With Juno's many instruments, a~fly-by can probe the nature of 3I/ATLAS far better than telescopes on~Earth.

We have further shown that much closer distances  to 3I/ATLAS ($\sim${27} million \si{km}) can be achieved with smaller $\Delta$V requirements should Juno have a relatively low level of propellant mass~remaining.

Small corrections to Juno's path might be needed if cometary activity of 3I/ATLAS will be intensified as it comes closer to the Sun and its non-gravitational acceleration will change its expected~trajectory.

\vspace{+6pt}
%%%%%%%%%%%%%%%%%%%%%%%%%%%%%%%%%%%%%%%%%%
\authorcontributions{{Conceptualization, A.C., A.L., A.H.; methodology, A.H.; software, A.H., validation, A.H., draft preparation, A.H.; investigation, A.H., A.L. A.C.; supervision, A.L., A.C.; review and editing A. L., A. H., All authors have read and agreed to the published version of the manuscript.}} 
%MDPI: For research articles with several authors, a short paragraph specifying their individual contributions must be provided. The following statements should be used ``Conceptualization, X.X. and Y.Y.; methodology, X.X.; software, X.X.; validation, X.X., Y.Y. and Z.Z.; formal analysis, X.X.; investigation, X.X.; resources, X.X.; data curation, X.X.; writing---original draft preparation, X.X.; writing---review and editing, X.X.; visualization, X.X.; supervision, X.X.; project administration, X.X.; funding acquisition, Y.Y. All authors have read and agreed to the published version of the manuscript.'', please turn to the  \href{http://img.mdpi.org/data/contributor-role-instruction.pdf}{CRediT taxonomy} for the term explanation. Authorship must be limited to those who have contributed substantially to the work~reported.}
% RESPONSE: DONE AS REQUIRED A.H.
\funding{{The APC was funded by Harvard Galileo Project, and Avi Loeb was supported in part by Harvard’s Black Hole Initiative and the Galileo Project. Adam Hibberd and Adam Crowl worked voluntarily for the Initiative for Interstellar Studies (i4is).}} %MDPI: Please add: ``This research received no external funding'' or ``This research was funded by NAME OF FUNDER grant number XXX.'' and  and ``The APC was funded by XXX''. Check carefully that the details given are accurate and use the standard spelling of funding agency names at \url{https://search.crossref.org/funding}, any errors may affect your future funding.}

\dataavailability{{Most data are contained within the article, for any additional data please refer to co-author Adam Hibberd.}} %MDPI: We encourage all authors of articles published in MDPI journals to share their research data. In this section, please provide details regarding where data supporting reported results can be found, including links to publicly archived datasets analyzed or generated during the study. Where no new data were created, or where data is unavailable due to privacy or ethical restrictions, a statement is still required. Suggested Data Availability Statements are available in section ``MDPI Research Data Policies'' at \url{https://www.mdpi.com/ethics}.} 

\acknowledgments{On 31 July 2025, Rep. Anna Paulina Luna sent a letter to NASA's {leadership} %MDPI: 1. Footnotes are not allowed. We have moved the content of this section to the main text. Please confirm. 2. Please add the access date (format: Date Month Year), e.g., accessed on 1 January 2020. Same as below.
{(\url{https://lweb.cfa.harvard.edu/\~loeb/APL_NASA.pdf}, accessed on 31 July 2025)},
% RESPONSE: DONE AS REQUIRED A.H.
urging a study of the amount of propellant left in Juno and repurposing it to probe 3I/ATLAS, based on this paper.
We thank Scott Bolton, Principal Investigator of the Juno mission, for~the helpful comments. Avi Loeb was supported in part by Harvard's Black Hole Initiative and the Galileo~Project.}

\conflictsofinterest{{The authors declare no conflicts of interest}} %MDPI: Declare conflicts of interest or state ``The authors declare no conflicts of interest.'' Authors must identify and declare any personal circumstances or interest that may be perceived as inappropriately influencing the representation or interpretation of reported research results. Any role of the funders in the design of the study; in the collection, analyses or interpretation of data; in the writing of the manuscript; or in the decision to publish the results must be declared in this section. If there is no role, please state ``The funders had no role in the design of the study; in the collection, analyses, or interpretation of data; in the writing of the manuscript; or in the decision to publish the results''.} 
% RESPONSE: DONE AS REQUIRED. A.H.
\begin{adjustwidth}{-\extralength}{0cm}
%\centering %% If there is a figure in wide page, please release command \centering
\reftitle{References}

% Please provide either the correct journal abbreviation (e.g., according to the “List of Title Word Abbreviations” http://www.issn.org/services/online-services/access-to-the-ltwa/) or the full name of the journal.

\bibliography{JUNO.bib}

\begin{thebibliography}{999}

\bibitem[Seligman et~al.(2025)Seligman, Micheli, Farnocchia, Denneau, Noonan,
  Hsieh, Santana-Ros, Tonry, Auchettl, Conversi, Devogèle, Faggioli,
  Feinstein, Fenucci, Ferrais, Frincke, Hainaut, Hart, Hoffman, Holt,
  Hoogendam, Huber, Jehin, Kareta, Keane, Kelley, Lister, Mandt, Marčeta,
  Meech, Miftah, Morgan, Ocaña, Peña-Asensio, Shappee, Siverd, Taylor,
  Tucker, Wainscoat, Weryk, Wray, Yaginuma, Yang, Ye, and
  Zhang]{seligman2025discovery}
Seligman, D.Z.; Micheli, M.; Farnocchia, D.; Denneau, L.; Noonan, J.W.; Hsieh,
  H.H.; Santana-Ros, T.; Tonry, J.; Auchettl, K.; Conversi, L.;  et~al.
\newblock Discovery and Preliminary Characterization of a Third Interstellar
  Object: 3I/ATLAS,  2025,  \href{http://arxiv.org/abs/2507.02757}{{\normalfont
  [arXiv:astro-ph.EP/2507.02757]}}.

\bibitem[Loeb(2025)]{Loeb_2025}
Loeb, A.
\newblock 3I/ATLAS is Smaller or Rarer than It Looks.
\newblock {\em Research Notes of the AAS} {\bf 2025}, {\em 9},~178.
\newblock {\url{https://doi.org/10.3847/2515-5172/adee06}}.

\bibitem[Bolin et~al.(2025)Bolin, Belyakov, Fremling, Graham, Abdelaziz,
  Elhosseiny, Gray, Ingebretsen, Jewett, Karpov, Kilic, Mašek, Molham,
  Roderick, Takey, Lisse, Abron, Coughlin, Hsieh, Noll, and Wong]{bolin2025}
Bolin, B.T.; Belyakov, M.; Fremling, C.; Graham, M.J.; Abdelaziz, A.M.;
  Elhosseiny, E.; Gray, C.L.; Ingebretsen, C.; Jewett, G.; Karpov, S.;  et~al.
\newblock Interstellar comet 3I/ATLAS: discovery and physical description,
  2025,  \href{http://arxiv.org/abs/2507.05252}{{\normalfont
  [arXiv:astro-ph.EP/2507.05252]}}.

\bibitem[Alvarez-Candal et~al.(2025)Alvarez-Candal, Rizos, Lara, Santos-Sanz,
  Gutierrez, Ortiz, and Morales]{alvarezcandal2025}
Alvarez-Candal, A.; Rizos, J.L.; Lara, L.M.; Santos-Sanz, P.; Gutierrez, P.J.;
  Ortiz, J.L.; Morales, N.
\newblock X-SHOOTER Spectrum of Comet C/2025 N1: Insights into a Distant
  Interstellar Visitor,  2025,
  \href{http://arxiv.org/abs/2507.07312}{{\normalfont
  [arXiv:astro-ph.EP/2507.07312]}}.

\bibitem[Opitom et~al.(2025)Opitom, Snodgrass, Jehin, Bannister, Bufanda, Deam,
  Dorsey, Ferrais, Hmiddouch, Knight, Kokotanekova, Leicester, Marsset, Murphy,
  Okoth, Ridden-Harper, Donckt, Ferellec, Hutsemekers, Lippi, Manfroid, and
  Benkhaldoun]{opitom2025}
Opitom, C.; Snodgrass, C.; Jehin, E.; Bannister, M.T.; Bufanda, E.; Deam, S.E.;
  Dorsey, R.; Ferrais, M.; Hmiddouch, S.; Knight, M.M.;  et~al.
\newblock Snapshot of a new interstellar comet: 3I/ATLAS has a red and
  featureless spectrum,  2025,
  \href{http://arxiv.org/abs/2507.05226}{{\normalfont
  [arXiv:astro-ph.EP/2507.05226]}}.

\bibitem[{Chandler} et~al.(2025){Chandler}, {Bernardinelli}, {Juri{\'c}},
  {Singh}, {Hsieh}, {Sullivan}, {Jones}, {Kurlander}, {Vavilov}, {Eggl},
  {Holman}, {Spoto}, {Schwamb}, {Christensen}, {Beebe}, {Roodman}, {Lim},
  {Jenness}, {Bosch}, {Smart}, {Bellm}, {MacBride}, {Rawls}, {Greenstreet},
  {Slater}, {Heinze}, {Ivezi{\'c}}, {Blum}, {Connolly}, {Daues}, {Makadia},
  {Gower}, {Bryce Kalmbach}, {Monet}, {Bannister}, {Dones}, {Dorsey}, {Fraser},
  {Forbes}, {Fuentes}, {Holt}, {Inno}, {Jones}, {Knight}, {Lintott}, {Lister},
  {Lupton}, {Mendoza Magbanua}, {Malhotra}, {Mueller}, {Murtagh}, {Pandey},
  {Reach}, {Samarasinha}, {Seligman}, {Snodgrass}, {Solontoi}, {Szab{\'o}},
  {White}, {Womack}, {Young}, {Allbery}, {Armellin}, {Aubourg}, {Avdellidou},
  {Azfar}, {Bauer}, {Bechtol}, {Belyakov}, {Benecchi}, {Bertini}, {Bolin},
  {Bose}, {Buchanan}, {Boucaud}, {Boufleur}, {Boutigny}, {Braga-Ribas},
  {Calabrese}, {Camargo}, {Caplar}, {Carry}, {Carvajal}, {Choi}, {Cowan},
  {Croft}, {{\'C}uk}, {Daruich}, {Daubard}, {Davenport}, {Daylan}, {Delgado},
  {Devillepoix}, {Doherty}, {Donaldson}, {Drass}, {Deppe}, {Dubois-Felsmann},
  {Economou}, {Eduardo}, {Farnocchia}, {Frissell}, {Fedorets}, {Fernandes},
  {Fulle}, {Gerdes}, {Gibbs}, {Gillan}, {Guy}, {Hammergren}, {Hanushevsky},
  {Hernandez}, {Hestroffer}, {Hopkins}, {Granvik}, {Ieva}, {Irving}, {Jannuzi},
  {Jimenez}, {Ramos Gomes-J{\'u}nior}, {Juramy}, {Kahn}, {Kannawadi}, {Kang},
  {Kryszczy{\'n}ska}, {Kotov}, {Koumjian}, {Krughoff}, {Lage}, {Lange},
  {Levine}, {Li}, {Licandro}, {Lin}, {Lust}, {Lyttle}, {Mahabal}, {Mahlke},
  {Plazas Malag{\'o}n}, {Salazar Manzano}, {Marc}, {Margoti}, {Mar{\v{c}}eta},
  {Menanteau}, {Meyers}, {Mills}, {Morato}, {More}, {Morrison}, {Moulane},
  {Mu{\~n}oz-Guti{\'e}rrez}, {Newcomer F.}, {O'Connor}, {Oldag}, {Oldroyd},
  {O'Mullane}, {Opitom}, {Oszkiewicz}, {Page}, {Patterson}, {Payne}, {Peloton},
  {Pereira}, {Peterson}, {Polin}, {Pollek}, {Polen}, {Qiu}, {Ragozzine},
  {Rajagopal}, {van Reeven}, {Rice}, {Ridgway}, {Rivkin}, {Robinson},
  {Ro{\.z}ek}, {Salnikov}, {S{\'a}nchez}, {Sarid}, {Schambeau}, {Scolnic},
  {Schindler}, {Seaman}, {Jacques}, {Shaw}, {Shugart}, {Sick}, {Siraj},
  {Sitarz}, {Sobhani}, {Soldahl}, {Stalder}, {Stetzler}, {Swinbank}, {Szigeti},
  {Tauraso}, {Thornton}, {Tonietti}, {Trilling}, and {Trujillo}]{Chandler2025}
{Chandler}, C.O.; {Bernardinelli}, P.H.; {Juri{\'c}}, M.; {Singh}, D.; {Hsieh},
  H.H.; {Sullivan}, I.; {Jones}, R.L.; {Kurlander}, J.A.; {Vavilov}, D.;
  {Eggl}, S.;  et~al.
\newblock {NSF-DOE Vera C. Rubin Observatory Observations of Interstellar Comet
  3I/ATLAS (C/2025 N1)}.
\newblock {\em arXiv e-prints} {\bf 2025}, p. arXiv:2507.13409,
  \href{http://arxiv.org/abs/2507.13409}{{\normalfont
  [arXiv:astro-ph.EP/2507.13409]}}.
\newblock {\url{https://doi.org/10.48550/arXiv.2507.13409}}.

\bibitem[{Belyakov} et~al.(2025){Belyakov}, {Fremling}, {Graham}, {Bolin},
  {Kilic}, {Jewett}, {Lisse}, {Ingebretsen}, {Davis}, and {Wong}]{Belyakov2025}
{Belyakov}, M.; {Fremling}, C.; {Graham}, M.J.; {Bolin}, B.T.; {Kilic}, M.;
  {Jewett}, G.; {Lisse}, C.M.; {Ingebretsen}, C.; {Davis}, M.R.; {Wong}, I.
\newblock {Palomar and Apache Point Spectrophotometry of Interstellar Comet
  3I/ATLAS}.
\newblock {\em Research Notes of the American Astronomical Society} {\bf 2025},
  {\em 9},~194.
\newblock {\url{https://doi.org/10.3847/2515-5172/adf059}}.

\bibitem[Flekk{\o}y et~al.(2019)Flekk{\o}y, Luu, and Toussaint]{Flekky2019}
Flekk{\o}y, E.G.; Luu, J.; Toussaint, R.
\newblock {The interstellar object'Oumuamua as a fractal dust aggregate}.
\newblock {\em The Astrophysical Journal Letters} {\bf 2019}, {\em 885},~L41.

\bibitem[Seligman and Laughlin(2020)]{Seligman2020}
Seligman, D.; Laughlin, G.
\newblock {Evidence that 1I/2017 U1 (Oumuamua) was composed of molecular
  hydrogen ice}.
\newblock {\em arXiv preprint arXiv:2005.12932} {\bf 2020}.

\bibitem[Jackson and Desch(2021)]{Jackson2021}
Jackson, A.P.; Desch, S.J.
\newblock 1I/‘Oumuamua as an N 2 Ice Fragment of an exo‐Pluto Surface: I.
  Size and Compositional Constraints.
\newblock {\em Journal of Geophysical Research: Planets} {\bf 2021}, {\em 126}.
\newblock {\url{https://doi.org/10.1029/2020je006706}}.

\bibitem[Desch and Jackson(2021)]{Desch2021}
Desch, S.J.; Jackson, A.P.
\newblock 1I/‘Oumuamua as an N 2 Ice Fragment of an Exo‐Pluto Surface II:
  Generation of N 2 Ice Fragments and the Origin of ‘Oumuamua.
\newblock {\em Journal of Geophysical Research: Planets} {\bf 2021}, {\em 126}.
\newblock {\url{https://doi.org/10.1029/2020je006807}}.

\bibitem[Bialy and Loeb(2018)]{Bialy2018}
Bialy, S.; Loeb, A.
\newblock {Could Solar Radiation Pressure Explain ‘Oumuamua's Peculiar
  Acceleration?}
\newblock {\em The Astrophysical Journal Letters} {\bf 2018}, {\em 868},~L1.

\bibitem[Raymond et~al.(2018)Raymond, Armitage, and Veras]{Raymond2018}
Raymond, S.N.; Armitage, P.J.; Veras, D.
\newblock Interstellar Object ’Oumuamua as an Extinct Fragment of an Ejected
  Cometary Planetesimal.
\newblock {\em The Astrophysical Journal} {\bf 2018}, {\em 856},~L7.
\newblock {\url{https://doi.org/10.3847/2041-8213/aab4f6}}.

\bibitem[Jewitt and Luu(2019)]{Jewitt_2019}
Jewitt, D.; Luu, J.
\newblock Initial Characterization of Interstellar Comet 2I/2019 Q4 (Borisov).
\newblock {\em The Astrophysical Journal Letters} {\bf 2019}, {\em 886},~L29.
\newblock {\url{https://doi.org/10.3847/2041-8213/ab530b}}.

\bibitem[{i4is}(2025)]{I4IS}
{i4is}.
\newblock {Initiative for Interstellar Studies},  2025.

\bibitem[{Hein} et~al.(2019){Hein}, {Perakis}, {Eubanks}, {Hibberd}, {Crowl},
  {Hayward}, {Kennedy}, and {Osborne}]{HPE19}
{Hein}, A.M.; {Perakis}, N.; {Eubanks}, T.M.; {Hibberd}, A.; {Crowl}, A.;
  {Hayward}, K.; {Kennedy}, R.G.; {Osborne}, R.
\newblock {Project Lyra: Sending a spacecraft to 1I/'Oumuamua (former A/2017
  U1), the interstellar asteroid}.
\newblock {\em Acta Astronaut.} {\bf 2019}, {\em 161},~552--561.
\newblock {\url{https://doi.org/10.1016/j.actaastro.2018.12.042}}.

\bibitem[{Hein} et~al.(2022){Hein}, {Eubanks}, {Lingam}, {Hibberd}, {Fries},
  {Schneider}, {Kervella}, {Kennedy}, {Perakis}, and {Dachwald}]{HEL22}
{Hein}, A.M.; {Eubanks}, T.M.; {Lingam}, M.; {Hibberd}, A.; {Fries}, D.;
  {Schneider}, J.; {Kervella}, P.; {Kennedy}, R.; {Perakis}, N.; {Dachwald}, B.
\newblock {Interstellar Now! Missions to Explore Nearby Interstellar Objects}.
\newblock {\em Adv. Space Res.} {\bf 2022}, {\em 69},~402--414.
\newblock {\url{https://doi.org/10.1016/j.asr.2021.06.052}}.

\bibitem[{Hibberd} et~al.(2020){Hibberd}, {Hein}, and {Eubanks}]{HHE20}
{Hibberd}, A.; {Hein}, A.M.; {Eubanks}, T.M.
\newblock {Project Lyra: Catching 1I/'Oumuamua - Mission opportunities after
  2024}.
\newblock {\em Acta Astronautica} {\bf 2020}, {\em 170},~136--144,
  \href{http://arxiv.org/abs/1902.04935}{{\normalfont
  [arXiv:physics.space-ph/1902.04935]}}.
\newblock {\url{https://doi.org/10.1016/j.actaastro.2020.01.018}}.

\bibitem[{Hibberd} and {Hein}(2021)]{HH21}
{Hibberd}, A.; {Hein}, A.M.
\newblock {Project Lyra: Catching 1I/'Oumuamua-Using Nuclear Thermal Rockets}.
\newblock {\em Acta Astronaut.} {\bf 2021}, {\em 179},~594--603,
  \href{http://arxiv.org/abs/2008.05435}{{\normalfont
  [arXiv:astro-ph.IM/2008.05435]}}.
\newblock {\url{https://doi.org/10.1016/j.actaastro.2020.11.038}}.

\bibitem[{Hibberd}(2023{\natexlab{a}})]{AH23}
{Hibberd}, A.
\newblock {Project Lyra: Another possible trajectory to 1I/’Oumuamua}.
\newblock {\em Acta Astronaut.} {\bf 2023}, {\em 211},~431--434.
\newblock {\url{https://doi.org/10.1016/j.actaastro.2023.06.029}}.

\bibitem[{Hibberd}(2023{\natexlab{b}})]{HA23}
{Hibberd}, A.
\newblock {Project Lyra: The Way to Go and the Launcher to Get There}.
\newblock {\em arXiv e-prints} {\bf 2023}, p. arXiv:2305.03065,
  \href{http://arxiv.org/abs/2305.03065}{{\normalfont
  [arXiv:astro-ph.IM/2305.03065]}}.
\newblock {\url{https://doi.org/10.48550/arXiv.2305.03065}}.

\bibitem[Seligman and Laughlin(2018)]{Seligman_2018}
Seligman, D.; Laughlin, G.
\newblock The Feasibility and Benefits of In Situ Exploration of
  ‘Oumuamua-like Objects.
\newblock {\em The Astronomical Journal} {\bf 2018}, {\em 155},~217.
\newblock {\url{https://doi.org/10.3847/1538-3881/aabd37}}.

\bibitem[{Hibberd} et~al.(2021){Hibberd}, {Perakis}, and {Hein}]{HPH21}
{Hibberd}, A.; {Perakis}, N.; {Hein}, A.M.
\newblock {Sending a spacecraft to interstellar comet 2I/Borisov}.
\newblock {\em Acta Astronaut.} {\bf 2021}, {\em 189},~584--592.
\newblock {\url{https://doi.org/10.1016/j.actaastro.2021.09.006}}.

\bibitem[Yaginuma et~al.(2025)Yaginuma, Frincke, Seligman, Mandt,
  DellaGiustina, Peña-Asensio, Taylor, and
  Nolan]{yaginuma2025feasibilityspacecraftflybyinterstellar}
Yaginuma, A.; Frincke, T.; Seligman, D.Z.; Mandt, K.; DellaGiustina, D.N.;
  Peña-Asensio, E.; Taylor, A.G.; Nolan, M.C.
\newblock The Feasibility of a Spacecraft Flyby with the Third Interstellar
  Object 3I/ATLAS from Earth or Mars,  2025,
  \href{http://arxiv.org/abs/2507.15755}{{\normalfont
  [arXiv:astro-ph.EP/2507.15755]}}.

\bibitem[Eubanks et~al.(2025)Eubanks, Bills, Hibberd, Blase, Hein, III,
  Coffinet, Schneider, Kervella, and
  de~Olea~Ballester]{eubanks20253iatlasc2025n1direct}
Eubanks, T.M.; Bills, B.G.; Hibberd, A.; Blase, W.P.; Hein, A.M.; III, R.G.K.;
  Coffinet, A.; Schneider, J.; Kervella, P.; de~Olea~Ballester, C.G.
\newblock 3I/ATLAS (C/2025 N1): Direct Spacecraft Exploration of a Possible
  Relic of Planetary Formation at "Cosmic Noon",  2025,
  \href{http://arxiv.org/abs/2508.15768}{{\normalfont
  [arXiv:astro-ph.EP/2508.15768]}}.

\bibitem[Conway and Paris(2010)]{CONWAY}
Conway, B.A.; Paris, S.W.
\newblock Spacecraft trajectory optimization using direct transcription and
  nonlinear programming.
\newblock {\em Spacecraft trajectory optimization} {\bf 2010}, {\em 29},~37.

\bibitem[Jezewski(1975)]{jezewski1975primer}
Jezewski, D.J.
\newblock Primer vector theory and applications.
\newblock Technical report,  1975.

\bibitem[Cage et~al.(1994)Cage, Kroo, and Braun]{cage1994}
Cage, P.; Kroo, I.; Braun, R.
\newblock Interplanetary trajectory optimization using a genetic algorithm.
\newblock In Proceedings of the Astrodynamics Conference,  1994, p. 3773.

\bibitem[Hibberd(2017)]{OITS_info}
Hibberd, A.
\newblock Github repository for OITS.,  2017.

\bibitem[{Hibberd}(2022)]{AH2}
{Hibberd}, A.
\newblock {Intermediate Points for Missions to Interstellar Objects Using
  Optimum Interplanetary Trajectory Software}.
\newblock {\em arXiv e-prints} {\bf 2022}, p. arXiv:2205.10220,
  \href{http://arxiv.org/abs/2205.10220}{{\normalfont
  [arXiv:astro-ph.EP/2205.10220]}}.

\bibitem[{Le Digabel}(2011)]{LeDigabel2011}
{Le Digabel}, S.
\newblock {Algorithm 909: NOMAD: Nonlinear optimization with the MADS
  algorithm}.
\newblock {\em ACM Transactions on Mathematical Software (TOMS)} {\bf 2011},
  {\em 37},~44.

\bibitem[Schlueter et~al.(2009)Schlueter, Egea, and
  Banga]{Schlueter_et_al_2009}
Schlueter, M.; Egea, J.; Banga, J.
\newblock Extended Ant Colony Optimization for non-convex Mixed Integer
  Nonlinear Programming.
\newblock {\em Computers and Operations Research} {\bf 2009}, {\em
  36},~2217--2229.
\newblock {\url{https://doi.org/10.1016/j.cor.2008.08.015}}.

\bibitem[Schlueter and Gerdts(2010)]{Schlueter_Gerdts_2010}
Schlueter, M.; Gerdts, M.
\newblock The Oracle Penalty Method.
\newblock {\em Journal of Global Optimization} {\bf 2010}, {\em 47},~293--325.
\newblock {\url{https://doi.org/10.1007/s10898-009-9477-0}}.

\bibitem[Schlueter et~al.(2013)Schlueter, Erb, Gerdts, Kemble, and
  Ruckmann]{Schlueter_et_al_2013}
Schlueter, M.; Erb, S.; Gerdts, M.; Kemble, S.; Ruckmann, J.
\newblock MIDACO on MINLP Space Applications.
\newblock {\em Advances in Space Research} {\bf 2013}, {\em 51},~1116--1131.
\newblock {\url{https://doi.org/10.1016/j.asr.2012.11.006}}.

\bibitem[{NAIF}(2025)]{NAIF}
{NAIF}.
\newblock {PLanetary Data System Navigation Node},  2025.

\bibitem[Blanco and Mungan(2021)]{Blanco2021}
Blanco, P.R.; Mungan, C.E.
\newblock High-speed escape from a circular orbit.
\newblock {\em American Journal of Physics} {\bf 2021}, {\em 89},~72--79,
  \href{http://arxiv.org/abs/https://pubs.aip.org/aapt/ajp/article-pdf/89/1/72/20094902/72\_1\_10.0001956.pdf}{{\normalfont
  [https://pubs.aip.org/aapt/ajp/article-pdf/89/1/72/20094902/72\_1\_10.0001956.pdf]}}.
\newblock {\url{https://doi.org/10.1119/10.0001956}}.

\bibitem[{Bate} et~al.(1971){Bate}, {Mueller}, and {White}]{Bate1971}
{Bate}, R.R.; {Mueller}, D.D.; {White}, J.E.
\newblock {\em {Fundamentals of astrodynamics}}; New York: Dover Publications,
  1971.

\bibitem[{NASA}(2011)]{JUNOPK}
{NASA}.
\newblock {Juno Launch Press Kit},  2011.

\bibitem[{Rein} and {Liu}(2012)]{2012A&A...537A.128R}
{Rein}, H.; {Liu}, S.F.
\newblock {REBOUND: an open-source multi-purpose N-body code for collisional
  dynamics}.
\newblock {\em Astronomy and Astrophysics} {\bf 2012}, {\em 537},~A128,
  \href{http://arxiv.org/abs/1110.4876}{{\normalfont
  [arXiv:astro-ph.EP/1110.4876]}}.
\newblock {\url{https://doi.org/10.1051/0004-6361/201118085}}.

\bibitem[{Rein} and {Spiegel}(2015)]{2015MNRAS.446.1424R}
{Rein}, H.; {Spiegel}, D.S.
\newblock {IAS15: a fast, adaptive, high-order integrator for gravitational
  dynamics, accurate to machine precision over a billion orbits}.
\newblock {\em Monthly Notes of the Royal Astronomical Society} {\bf 2015},
  {\em 446},~1424--1437,  \href{http://arxiv.org/abs/1409.4779}{{\normalfont
  [arXiv:astro-ph.EP/1409.4779]}}.
\newblock {\url{https://doi.org/10.1093/mnras/stu2164}}.

\end{thebibliography}

%%%%%%%%%%%%%%%%%%%%%%%%%%%%%%%%%%%%%%%%%%
%% for journal Sci
%\reviewreports{\\
%Reviewer 1 comments and authors’ response\\
%Reviewer 2 comments and authors’ response\\
%Reviewer 3 comments and authors’ response
%}
%%%%%%%%%%%%%%%%%%%%%%%%%%%%%%%%%%%%%%%%%%
\PublishersNote{}
\end{adjustwidth}

\end{document}